\documentclass[10pt,twocolumn,letterpaper]{article}

\usepackage{cvpr}
\usepackage{times}
\usepackage{epsfig}
\usepackage{graphicx}
\usepackage{amsmath}
\usepackage{amssymb}

\usepackage{multirow}
\usepackage{bigstrut}
\usepackage{booktabs}
\usepackage{color}
\usepackage{fancyhdr}

\usepackage[pagebackref=true,breaklinks=true,letterpaper=true,colorlinks,bookmarks=false]{hyperref}

\cvprfinalcopy 

\def\cvprPaperID{6293} 

\begin{document}
\title{Supervised Raw Video Denoising with a Benchmark Dataset on Dynamic Scenes}

\author{
Huanjing Yue \quad Cong Cao \quad Lei Liao \quad Ronghe Chu \quad Jingyu Yang\thanks{This work was supported in part by the National Natural Science Foundation of China under Grant 61672378, Grant 61771339, and Grant 61520106002. \textit{Corresponding author: Jingyu Yang.}}
\\
School of Electrical and Information Engineering, Tianjin University, Tianjin, China\\
{\tt\small \{huanjing.yue, caocong\_123, leolei, chu\_rh, yjy\}@tju.edu.cn}\\
{\tt\small \url{https://github.com/cao-cong/RViDeNet}}
}

\maketitle

\begin{abstract}

In recent years, the supervised learning strategy for real noisy image denoising has been emerging and has achieved
promising results. In contrast, realistic noise removal for raw noisy videos is rarely studied due to the lack of
noisy-clean pairs for dynamic scenes. Clean video frames for dynamic scenes cannot be captured with a long-exposure  shutter or averaging multi-shots as was done for static images. In this paper, we solve this problem by creating motions for controllable objects, such as toys, and capturing each static moment for multiple times to generate clean video frames. In this way, we construct a dataset with 55 groups of noisy-clean videos with ISO values ranging from 1600 to 25600. To our knowledge, this is the first dynamic video dataset with noisy-clean pairs. Correspondingly, we propose a raw video denoising network (RViDeNet) by exploring the temporal, spatial, and channel correlations of video frames. Since the raw video has Bayer patterns, we pack it into four sub-sequences, i.e RGBG sequences, which are denoised by the proposed RViDeNet separately and finally fused into a clean video. In addition, our network not only outputs a raw denoising result,
but also the sRGB result by going through an image signal processing (ISP) module, which enables users to generate the sRGB result with their favourite ISPs. Experimental results demonstrate that our method outperforms state-of-the-art video and raw image denoising algorithms on both indoor and outdoor videos.

\end{abstract}

\section{Introduction}

Capturing videos under low-light conditions with  high ISO settings would inevitably introduce much noise \cite{chen2019seeing}, which dramatically deteriorates the visual quality and affects the followed analysis of these videos. Therefore, video denoising is essential in improving the quality of low-light videos.

However, due to the non-linear image signal processing (ISP), such as demosaicing, white balancing and color correction, the noise in the sRGB domain is more complex than Gaussian noise \cite{nam2016holistic}. Therefore, Gaussian noise removal methods cannot be directly used for realistic noise removal \cite{xu2018external,xu2017multi,xu2018trilateral}. On the other hand, convolutional neural networks (CNNs) enable us to learn the complex mapping between the noisy image and the clean image. Therefore, many CNN based realistic noise removal methods have emerged in recent years \cite{anwar2019real,guo2019toward, yue2019high}. These methods usually first build noisy-clean image pairs, in which the noisy image is captured with short exposure under high ISO mode and the clean image is the average of multiple noisy images of the same scene. Then, they design sophisticated networks to learn the mapping between the noisy image and clean image. Since this kind of image pairs are tedious to prepare, some methods propose to utilize both synthesized and real data to train the network \cite{guo2019toward, chen2018image}.

In contrast, the noise statistics in the raw domain, i.e. the direct readings from the image sensor, are simpler than these in the sRGB domain. In addition, the raw data contains the most original information since it was not affected by the following ISP. Therefore, directly performing denoising on the raw data is appealing. Correspondingly, there are many datasets built for raw image denoising by capturing the short-exposure raw noisy images and the long-exposure clean raw images \cite{abdelhamed2018high, plotz2017benchmarking, anaya2014renoir, chen2019learning}. However, there is still no dataset built for noisy and clean videos in the raw format since we cannot record the dynamic scenes without blurring using the long-exposure mode or averaging multiple shots of the moment. Therefore, many methods are proposed for raw image denoising, but raw video denoising is lagging behind. Very recently, Chen \textit{et al.} \cite{chen2019seeing} proposed to perform raw video denoising by capturing a dataset with static noisy and clean image sequences, and directly map the raw input to the sRGB output by simultaneously learning the noise removal and ISP. Nevertheless, utilizing static sequences to train the video enhancement network does not take advantage of the temporal correlations between neighboring frames and it relies on the well developed video denoising scheme VBM4D \cite{matteo2011video} to remove noise.

Based on the above observations, we propose to conduct video denoising in the raw domain and correspondingly construct a dataset with noisy-clean frames for dynamic scenes. There are mainly three contributions in this work.

First, we construct a benchmark dataset for supervised raw video denoising. In order to capture the moment for multiple times, we manually create movements for objects. For each moment, the noisy frame is captured under a high ISO mode, and the corresponding clean frame is obtained via averaging multiple noisy frames. In this way, we capture 55 groups of dynamic noisy-clean videos with ISO values ranging from 1600 to 25600. This dataset not only enables us to take advantage of the temporal correlations in denoising, but also enables the quantitative evaluation for real noisy videos.

Second, we propose an efficient raw video denoising network (RViDeNet) via exploring non-local spatial, channel, and temporal correlations. Since the noisy input is characterized by Bayer patterns, we split it into four separated sequences, i.e. RGBG sequences, and they go through the pre-denoising, alignment, non-local attention, and temporal fusion modules separately, and then reconstruct the noise-free version by spatial fusion.

Third, our network not only outputs the raw denoising result, but also the RGB result by going through an ISP module. In this way, our method enables users to adaptively generate the sRGB results with the ISP they prefer. Experimental results demonstrate that our method outperforms state-of-the-art video denoising and raw image denoising algorithms in both raw and sRGB domains on captured indoor and outdoor videos.


\section{Related Work}

In this section, we give a brief review of related work on video denoising, image and video processing with raw data, and noisy image and video datasets.

\subsection{Video Denoising}

%
%
%
%

In the literature, most video denoising methods are designed for Gaussian noise removal \cite{matteo2011video,ji2010robust,buades2016patch}. Among them, VBM4D is the benchmark denoising method \cite{matteo2011video}. Recently, deep learning based video denoising methods are emerging.
Chen \textit{et al.} \cite{chen2016deep} first proposed to apply recurrent neural network on video denoising in the sRGB domain. However, the performance is under the benchmark denoising method VBM4D. Hereafter, Xue \textit{et al.} \cite{xue2019video} proposed a task-oriented flow (ToF) to align frames via CNN and then performed the following denoising task. The recently proposed ViDeNN \cite{claus2019videnn} performs spatial denoising and temporal denoising sequentially and achieves better results than VBM4D. Tassano \textit{et al.} proposed DVDNet \cite{tassano2019dvdnet} and its fast version, called FastDVDnet \cite{tassano2019fastdvdnet} without explicit motion estimation, to deal with Gaussian noise removal with low computing complexity.

However, these methods are usually designed for Gaussian or synthesized noise removal, without considering the complex real noise produced in low-light capturing conditions. To our knowledge, only the work in \cite{chen2019seeing} deals with  realistic noise removal for videos. However, their training database contains only static sequences, which is inefficient in exploring temporal correlations of dynamic sequences. In this work, we construct a dynamic noisy video dataset, and correspondingly propose a RViDeNet to fully take advantage of the spatial, channel, and temporal correlations.

\subsection{Image and Video Processing with Raw Data}
Since visual information goes through the complex ISP to generate the final sRGB image, images in the raw domain contain the most visual information and the noise is simpler than that in the sRGB domain. Therefore, many works are proposed to process images processing in the raw domain.

With several constructed raw image denoising datasets \cite{anaya2014renoir,abdelhamed2018high,plotz2017benchmarking,chen2019learning}, raw image denoising methods have attracted much attention \cite{gharbi2016deep,chen2019learning}. Besides these datasets, Brooks \textit{et al.} \cite{brooks2018unprocessing} proposed an effective method to unprocess sRGB images back to the raw images, and achieved promising denoising performance on the DND dataset.
The winner of NTIRE 2019 Real Image Denoising Challenge proposed a Bayer preserving augmentation method for raw image denoising, and achieved state-of-the-art denoising results \cite{liu2019learning}.
Besides denoising, the raw sensor data has also been used in other image restoration tasks, such as image super-resolution \cite{xu2019towards,zhang2019zoom}, joint restoration and enhancement \cite{ratnasingam2019deep,schwartz2018deepisp,liang2019cameranet}. These works also demonstrate that directly processing the raw images can generate more appealing results than processing the sRGB images.


However, videos are rarely processed in the raw domain. Very recently, Chen \textit{et al.} \cite{chen2019seeing} proposed to perform video denoising by mapping raw frames to the sRGB ones with static frames as training data. Different from it, we propose to train a RViDeNet by mapping the raw data to both raw and sRGB outputs, which can generate flexible results for different users.

\subsection{Noisy Image and Video Datasets}
Since the training data is essential for realistic noise removal, many works focus on noisy-clean image pairs construction.
There are two strategies to generate clean images. One approach is generating the noise-free image by averaging multiple frames for one static scene and all the images are captured by a stationary camera with fixed settings \cite{nam2016holistic,yue2019high,xu2018real,abdelhamed2018high}. In this way, the clean image has similar brightness with the noisy ones. The noisy images in \cite{nam2016holistic,yue2019high,xu2018real} are saved in sRGB format.
Another strategy is capturing a static scene under low/high ISO setting and use the low ISO image as the ground truth of the noisy high ISO image, such as the RENOIR dataset \cite{anaya2014renoir}, the DND dataset \cite{plotz2017benchmarking}, and SID dataset \cite{chen2019learning}.
The images in RENOIR, DND, SIDD \cite{abdelhamed2018high}, and SID are all captured in raw format, and the sRGB images are synthesized according to some image ISP modules. Recently, the work in \cite{chen2019seeing} constructed a noisy-clean datasets for static scenes, where a clean frame corresponds to multiple noisy frames.

To our knowledge, there is still no noisy-clean video datasets since it is impossible to capture the dynamic scenes with long-exposure or multiple shots without introducing blurring artifacts. In this work, we solve this problem by manually create motions for objects. In this way, we can capture each motion for multiple times and produce the clean frame by averaging these shots.

\section{Raw Video Dataset}

\subsection{Captured Raw Video Dataset}
Since there is no realistic noisy-clean video dataset, we collected a raw video denoising dataset to facilitate related research. We utilized a surveillance camera with the sensor IMX385, which is able to continuously capture 20 raw frames per second.
The resolution for the Bayer image is $1920\times1080$.

\begin{figure}[htb]
    \centering
    \includegraphics[width=0.9\linewidth]{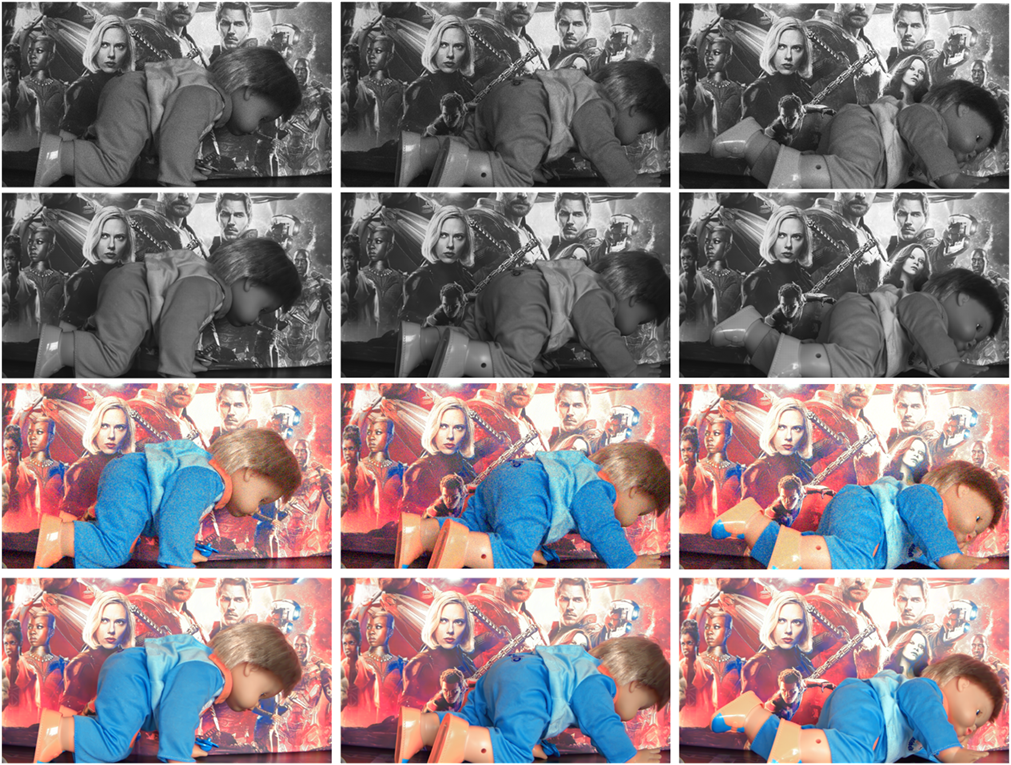}
    \caption{Sample frames of the captured noisy-clean video under ISO 25600. From left to right, they are respectively the 2nd, 3rd, and 4th  frames in the video. From top to down, each row lists the raw noisy video, raw clean video, sRGB noisy video, and sRGB clean video, respectively. The color videos are generated from raw video using our pre-trained ISP module.}
    \label{fig:dataset}
\end{figure}

The biggest challenge is how to simultaneously capture noisy videos and the corresponding clean ones for dynamic scenes. Capturing clean dynamic videos using low ISO and high exposure time will cause motion blur. To solve this problem, we propose to capture controllable objects, such as toys, and manually make motions for them. For each motion, we continuously captured $M$ noisy frames. The averaging of the $M$ frames is the ground truth (GT) noise-free frame. We do not utilize long exposure to capture the GT noise free frame since it will make the GT frame and noisy frames have different brightness.
Then, we moved the object and kept it still again to capture the next noisy-clean paired frame. Finally, we grouped all the single frames together according to their temporal order to generate the noisy video and its corresponding clean video.
We totally captured 11 different indoor scenes under 5 different ISO levels ranging from 1600 to 25600. Different ISO settings is used to capture different level noise. For each video, we captured seven frames. Fig. \ref{fig:dataset} presents the second, third, and forth frames of an captured video under ISO 25600. It can be observed that this video records the crawling motion of the doll.

Our camera is fixed to a tripod when capturing the continuous $M$ frames and therefore the captured frames are well aligned. Since higher ISO will introduce more noise, we captured 500 frames for the averaging when ISO is 25600. We note that there is still slight noise after averaging noisy frames, and we further applied BM3D \cite{dabov2007image} to the averaged frame to get a totally clean ground truth. The detailed information for our captured noisy-clean dataset is listed in the supplementary  material. These captured noisy-clean videos not only enable supervised training but also enable quantitative evaluation.

Since it is difficult to control outdoor objects, the above noisy-clean video capturing approach is only applied to indoor scenes. The captured 11 indoor scenes are split into training and validation set (6 scenes), and testing set (5 scenes).
We used the training set to finetune our model which has been pretrained on synthetic raw video dataset (detailed in the following section) and used the testing set to test our model.
We also captured another 50 outdoor dynamic videos under different ISO levels to further test our trained model.

\begin{figure*}
    \centering
    \includegraphics[width=0.9\linewidth]{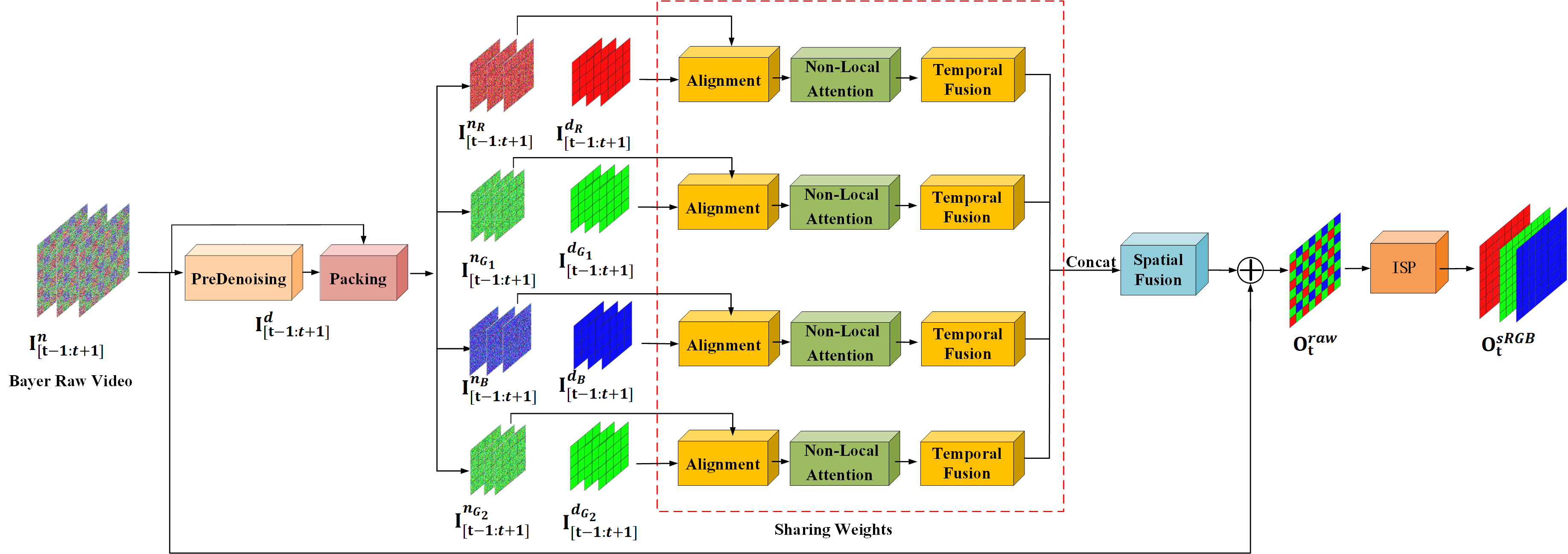}
    \caption{The framework of proposed RViDeNet. The input noisy sequence is packed into four sub-sequences according to the Bayer pattern and then go through alignment, non-local attention and temporal fusion modules separately, and finally fuse into a clean frame by spatial fusion. With the followed ISP module, a denoising result in the sRGB domain is also produced.}
    \label{fig:framework}
\end{figure*}

\subsection{Synthesized Raw Video Dataset}\label{Sec:syn}
Since it is difficult to capture videos for various moving objects, we further propose to synthesize noisy videos as supplementary training data. We choose four videos from MOTChallenge dataset \cite{MilanMOT16}, which contains scene motion, camera motion, or both. These videos are sRGB videos and each video has several hundreds of frames. We first utilize the image unprocessing method proposed in \cite{brooks2018unprocessing} to convert these sRGB videos to raw videos, which serve as the ground truth clean videos. Then, we add noise to create the corresponding noisy raw videos.

As demonstrated in \cite{MildenhallBurst, FoiPractical}, the noise in raw domain contains the shot noise modeled by Poisson noise and read noise modeled by Gaussian noise. This process is formulated as
\begin{equation}
x_p \sim \sigma_s^2\mathcal{P}(y_p/\sigma_s^2)+\mathcal{N}(0,\sigma_r^2)
\end{equation}
where $x_p$ is the noisy observation, $y_p$ is the true intensity at pixel $p$.  $\sigma_r$ and $\sigma_s$ are parameters for read and shot noise, which vary across images as sensor gain (ISO) changes. The first term represents the Poisson distribution with mean $y_p$ and variance $\sigma_s^2y_p$.  The second term represents Gaussian distribution with zero mean and variance $\sigma_r^2$.

Different from \cite{MildenhallBurst}, we calibrate the noise parameters for given cameras by capturing flat-field frames\footnote{{https://en.wikipedia.org/wiki/Flat-field\_correction}} and bias frames\footnote{{https://en.wikipedia.org/wiki/Bias\_frame}}. Flat-field frames are the images captured when sensor is uniformly illuminated. Rather than capturing many frames to estimate $\sigma_s$, which is the strategy used in \cite{FoiNoise}, capturing flat-field frames is faster.
Tuning camera to a specific ISO, we only need take images of a white paper on a uniformly lit wall under different exposure times. Then we compute estimated signal intensity against the corrected variance to determine $\sigma_s$. Bias frames are the images captured under a totally dark environment. Since there is no shot noise in bias frames, we use them to estimate $\sigma_r$\footnote{The technical details can be found in the supplementary material.}.


\section{The Proposed Method}
Given a set of consecutive frames (three frames in this work), we aim to recover the middle frame by exploring the spatial correlations inside the middle frame and the temporal correlations across neighboring frames. Fig. \ref{fig:framework} presents the framework of the proposed RViDeNet.

Since the captured raw frame is characterized by Bayer patterns, i.e. the color filter array pattern, we propose to split each raw frame into four sub-frames to make neighboring pixels be the filtered results of the same color filter (as shown in Fig. \ref{fig:framework}).
Inspired by the work of video restoration in \cite{wang2019edvr}, we utilize deformable convolutions \cite{dai2017deformable} to align the input frames instead of using the explicit flow information as done in \cite{xue2019video}. Then, we fuse the aligned features in temporal domain. Finally, we utilize the spatial fusion module to reconstruct the raw result. After the ISP module, we can obtain the sRGB output. In the following, we give details for these modules.

\subsection{PreDenoising and Packing}
As demonstrated in \cite{chen2019seeing}, the noise will heavily disturb the prediction of dense correspondences, which are the key module of many burst image denoising methods \cite{mildenhall2018burst,godard2018deep}, for videos. However, we find that using well-designed predenoising module can enable us to estimate the dense correspondences.

In this work, we train a single-frame based denoising network, i.e. the U-Net \cite{ronneberger2015u}, with synthesized raw noisy-clean image pairs to serve as the pre-denoising module. We use 230 clean raw images from SID \cite{chen2019learning} dataset, and synthesize noise using the method described in Sec. \ref{Sec:syn} to create noisy-clean pairs. Note that, pixels of different color channels in an raw image are mosaiced according to the Bayer pattern, i.e. the most similar pixels for each pixel are not its nearest neighbors, but are its secondary nearest neighbors. Therefore, we propose to pack the noisy frame $I_t^n$ into four channels, i.e. RGBG channels, to make spatially neighboring pixels have similar intensities. Then, these packed sub-frames go through the U-Net and the inverse packing process to generate the predenoising result, i.e. $I_t^d$.

For video denoising, our input is 2$N$+1 consecutive frames, i.e. $I^n_{[t-N:t+N]}$.  We extract the RGBG-sub-frames from each full-resolution frame. Then we concatenate all the sub-frames of each channel to form a sub sequence. In this way, we obtain four noisy sequences and four denoised sequences, and they are used in the alignment module. In the following, without specific clarifications, we still utilize $I^n_{[t-N:t+N]}$ to represent the reassembled sequences $I^{n_R}_{[t-N:t+N]}$, $I^{n_{G_1}}_{[t-N:t+N]}$ $I^{n_B}_{[t-N:t+N]}$, and $I^{n_{G_2}}_{[t-N:t+N]}$ for simplicity, since the following operations are the same for the four sequences.

\subsection{Alignment}
The alignment module aims at aligning the features of neighboring frames, i.e. the $(t+i)$-th frame, to that of the central frame, i.e. the $t$-th frame, which is realized by the deformable convolution \cite{dai2017deformable}.
For a deformable convolution kernel with $k$ locations, we utilize $w_k$ and $\bold{p}_k$  to represent the weight and pre-specified offset for the $k$-th location. The aligned features $\hat{F}^n_{t+i}$ at position $\bold{p}_0$ can be obtained by
\begin{equation}
\hat{F}^n_{t+i} (\bold{p}_0)= \sum_{k=1}^{K}w_k\cdot F^n_{t+i}(\bold{p}_0+\bold{p}_k+\triangle \bold{p}_k)\cdot \triangle m_k,
\label{eq:Dconv}
\end{equation}
where $F^n_{t+i}$ is the features extracted from the noisy image $I^n_{t+i}$. Since the noise will disturb the offsets estimation process, we utilize the denoised version to estimate the offsets. Namely, the learnable offset $\triangle \bold{p}_k$ and the modulation scalar $\triangle m_k$ are predicted from the concatenated features $[F^d_{t+i},F^d_t]$ via a network constructed by several convolution layers, i.e
\begin{equation}
\{\triangle \bold{p}\}_{t+i} = f([F^d_{t+i},F^d_t]),
\end{equation}
where $f$ is the mapping function, and $F^d_t$ is the features extracted from the denoised image $I^d_t$. For simplicity, we ignore the calculation process of $\triangle m_k$ in figures and descriptions.

\begin{figure}
    \centering
    \includegraphics[width=0.9\linewidth]{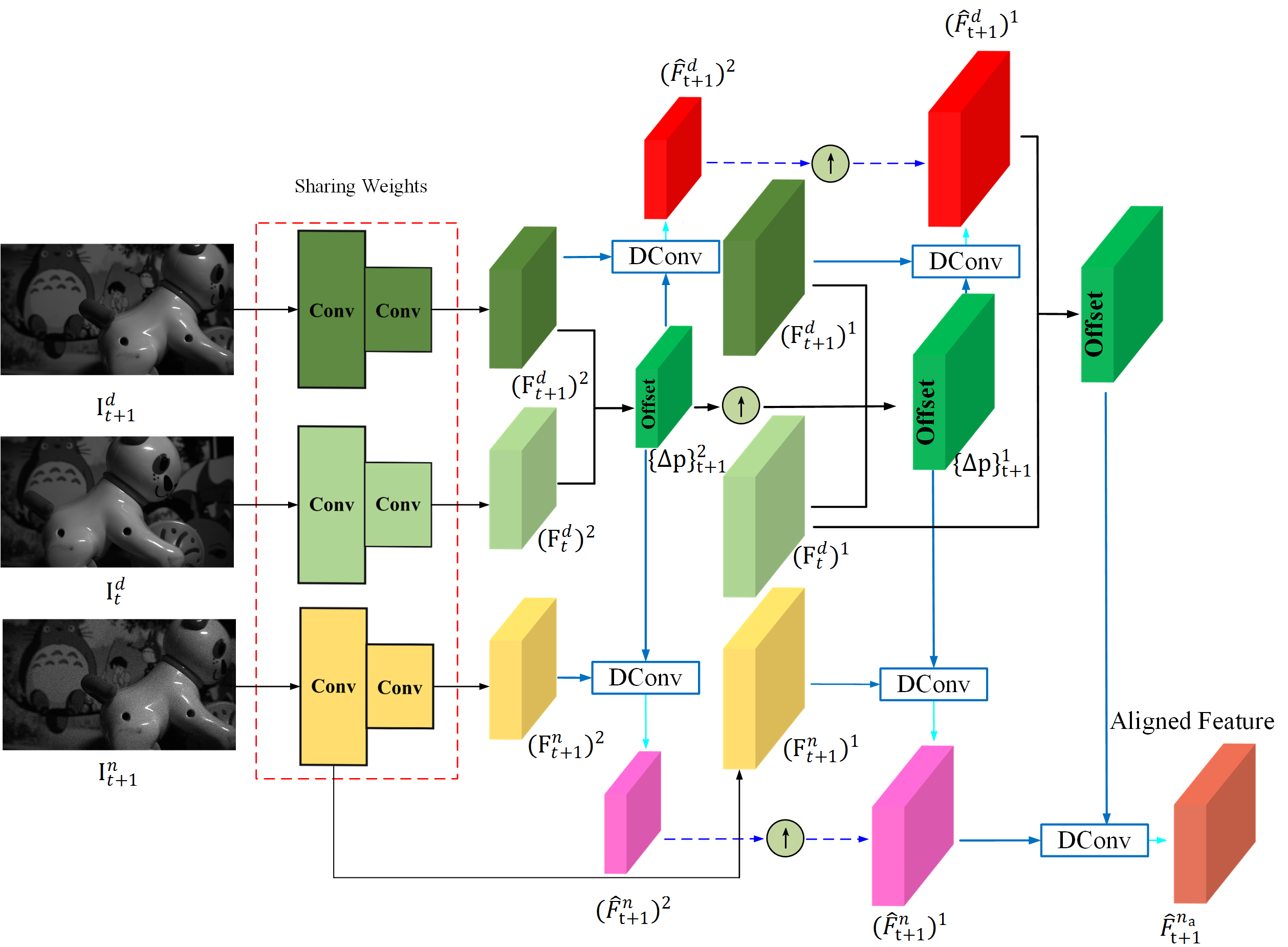}
    \caption{The pre-denoiseing result guided noisy frame alignment module. For simplicity, we only present the pyramidal processing with two levels. The feature extraction processes share weights.}
    \label{fig:alignment}
    \vspace{-0.1cm}
\end{figure}

Similar to \cite{wang2019edvr}, we utilize pyramidal processing and cascading refinement to deal with large movements. In this paper, we utilize three level pyramidal processing. For simplicity, Fig. \ref{fig:alignment} presents the pyramidal processing with only two levels. The features $(F^d_{t+1}$, $F^d_t)$ and $(F^n_{t+1}$, $F^n_t)$ are downsampled via strided convolution with a step size of 2 for $L$ times to form $L$-level pyramids of features.  Then, the offsets are calculated from the $l^{th}$ level, and the offsets are upsampled to the next $(l-1)^{th}$  level. The offsets in the $l^{th}$ level are calculated from both the upsampled offsets and the $l^{th}$ features. This process is denoted by
\begin{equation}
\{\triangle \bold{p}\}^l_{t+i} = f([(F^d_{t+i})^l,(F^d_t)^l],(\{\triangle \bold{p}\}^{l+1}_{t+i})^{\uparrow2}).
\end{equation}
Correspondingly, the aligned features for the noisy input and denoised input are obtained via
\begin{equation}
\begin{split}
(\hat{F}^n_{t+i})^l &= g(\text{DConv}((F^n_{t+i})^l,\{\triangle \bold{p}\}^l_{t+i}),((\hat{F}^n_{t+i})^{l+1})^{\uparrow2}),\\
(\hat{F}^d_{t+i})^l &= g(\text{DConv}((F^d_{t+i})^l,\{\triangle \bold{p}\}^l_{t+i}),((\hat{F}^d_{t+i})^{l+1})^{\uparrow2}),
\end{split}
\end{equation}
where DConv is the deformable convolution described in Eq. \ref{eq:Dconv} and $g$ is the mapping function realized by several convolution layers. After L levels alignment, $(\hat{F}^n_{t+i})^1$ is further refined by utilizing the offset calculated between $(\hat{F}_{t+i}^d)^1$ and $(F_t^d)^1$, and produce the final alignment result $\hat{F}^{n_a}_{t+i}$.

After the alignment for the two neighboring frames, we obtain $T\times C\times H\times W$ features, which contain the original central frame features extracted from $I_t^n$, and the aligned features from $I_{t+1}^n$ and $I_{t-1}^n$.

\subsection{Non-local Attention}
\begin{figure}
    \centering
    \includegraphics[width=0.9\linewidth]{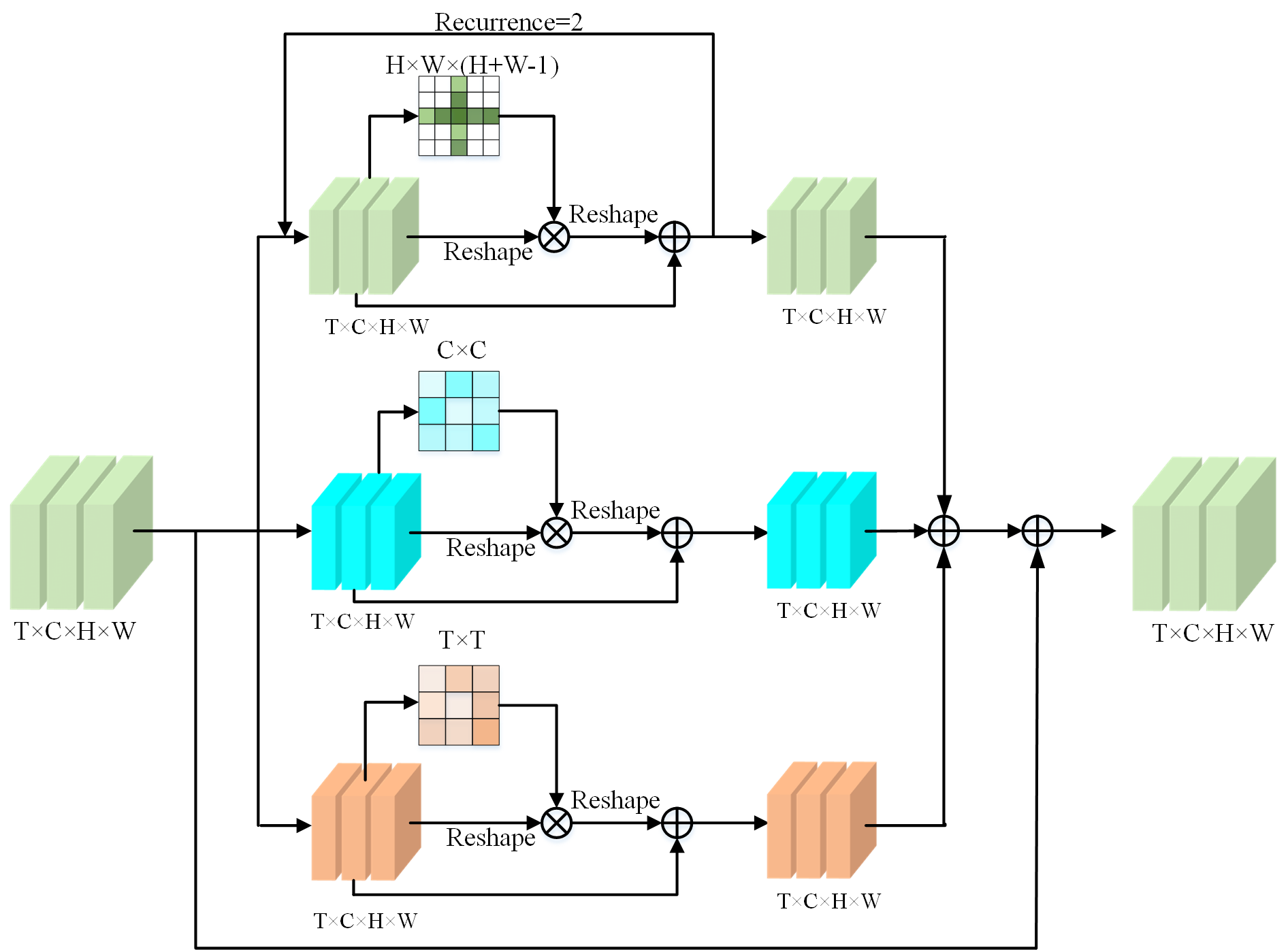}
    \caption{The non-local attention module. The green, blue, and orange modules represent the spatial, channel, and temporal attention respectively. }
    \label{fig:nonlocal}
      \vspace{-0.5cm}
\end{figure}

The DConv based alignment is actually the aggregation of the non-local similar features. To further enhance the aggregating process, we propose to utilize non-local attention module \cite{huang2019ccnet,fu2019dual,wang2018non}, which is widely used in semantic segmentation, to strengthen feature representations. Since 3D non local attention consumes huge costs, we utilize the separated attention modules \cite{fu2019dual}. Specifically, we utilize spatial attention, channel attention, and temporal attention to aggregate the long-range features. Then, the spatial, channel, and temporal enhanced features are fused together via element-wise summation. The original input is also added via residual connection. Note that, to reduce the computing and memory cost, we utilize criss-cross attention \cite{huang2019ccnet} to realize the spatial attention. This module is illustrated in Fig. \ref{fig:nonlocal}.

\subsection{Temporal Fusion}
Even though we have aligned the neighboring frame features with the central frame, these aligned neighboring frames still contribute differently to the denoising of the central frame due to the occlusions and alignment errors. Therefore, we adopt the element-wise temporal fusion strategy proposed in \cite{wang2019edvr} to adaptively fuse these features. The temporal similarities between the features of neighboring frames are calculated via dot product of features at the same position. Then the similarity is restricted to $[0,1]$ by the sigmoid function.  Hereafter, the features are weighted by  element-wise multiplication with the similarities, producing the weighted features $\tilde{F}^n_{t+i}$, i.e.
\begin{equation}
\tilde{F}^n_{t+i} = \hat{F}^{n_a}_{t+i}\odot S(\hat{F}^{n_a}_{t+i},\hat{F}^{n_a}_{t}),
\end{equation}
where $\odot$ represents the element-wise multiplication, $S$ represents the calculated similarity map, and $\hat{F}^{n_a}_{t}$ is the aligned features of frame $t$ after non-local attention.

An extra convolution layers is utilized to aggregate these concatenated weighted features, which are further weighted by spatial attentions by pyramidal processing \cite{wang2019edvr}. After temporal fusion, the features are squeezed to $1\times C\times H\times W$ again.

\subsection{Spatial Fusion}

After temporal fusion for the four sub-frame sequences, we utilize spatial fusion to fuse the four sequences together to generate a full-resolution output. The features $F^R_{\text{fus}},F^{G_1}_{\text{fus}},F^{B}_{\text{fus}}$, and $F^{G_2}_{\text{fus}}$ from the temporal fusion modules are concatenated together and then go through the spatial fusion network. The spatial fusion network is constructed by 10 residual blocks, a CBAM \cite{woocbam} module to enhance the feature representations, and a convolution layer to predict the noise with size $4\times H\times W$. Except the last output convolution layer, all the other convolution layer has $4\times C$ output channels.
Hereafter, the estimated noise in the four channels are reassembled into the full-resolution Bayer image via the inverse packing process. Finally, by adding the estimated noise with the original noisy input $I^n_{t}$, we obtain the raw denoising result $O^{\text{raw}}_{t}$ with size $1\times 2H \times 2W$.

\subsection{Image Signal Processing (ISP)}

We further pre-train the U-Net \cite{ronneberger2015u} as an ISP model to transfer $O^{\text{raw}}_{t}$ to the sRGB image $O^{\text{RGB}}_{t}$. We select 230 clean raw and sRGB pairs from SID dataset \cite{chen2019learning} to train the ISP model. By changing the training pairs, we can simulate  ISP of different cameras. In addition, ISP module can also be replaced by traditional ISP pipelines, such as DCRaw \footnote {https://dcraw.en.softonic.com/} and Adobe Camera Raw \footnote{https://helpx.adobe.com/camera-raw/using/supported-cameras.html}. Generating both the raw and sRGB outputs gives users more flexibility to choose images they prefer.

\subsection{Loss Functions}

Our loss function is composed by reconstruction loss and temporal consistent loss. The reconstruction loss constrains the restored image in both raw and sRGB domain to be similar with the ground truth. For temporal consistent loss, inspired by  \cite{chen2019seeing}, we choose four different noisy images for $I_t$ and utilize the first three frames to generate the denoising result $\hat{O}^{\text{raw}_1}_t$, and then utilize the last three frames to generate the denoising result $\hat{O}^{\text{raw}_2}_t$. Since $\hat{O}^{\text{raw}_1}_t$  and $\hat{O}^{\text{raw}_2}_t$ correspond to the same clean frame $I^{\text{raw}}_t$, we constrain them to be similar with each other and similar with $I^{\text{raw}}_t$. Different from  \cite{chen2019seeing}, we directly perform the loss functions in pixel domain other than the VGG feature domain. Our loss function is formulated as
\begin{equation}
\begin{split}
\mathcal{L}= &\mathcal{L}_{\text{rec}}+\lambda \mathcal{L}_{\text{tmp}},\\
\mathcal{L}_{\text{rec}}= &\|I^{\text{raw}}_t-O^{\text{raw}}_t\|_1+ \beta \|I^{\text{sRGB}}_t-O^{\text{sRGB}}_t\|_1,\\
\mathcal{L}_{\text{tmp}}= &\|\hat{O}^{\text{raw}_1}_t-\hat{O}^{\text{raw}_2}_t\|_1,\\
& + \gamma(\|I^{\text{raw}}_t-\hat{O}^{\text{raw}_1}_t\|_1+\|I^{\text{raw}}_t-\hat{O}^{\text{raw}_2}_t\|_1),
\label{mask loss}
\end{split}
\end{equation}
where $O^{\text{raw}}_t$ ($O^{\text{sRGB}}_t$) is the $t^{th}$ denoising frame in the raw (sRGB) domain for the consecutive noisy input $[I^n_{t-1},I^n_{t},I^n_{t+1}]$. $\lambda$, $\beta$, and $\gamma$ are the weighting parameters. At the training stage, our network is first trained with synthetic noisy sequences. We disable the temporal consistent loss by setting $\lambda=0$ and $\beta=0$ since minimizing $\mathcal{L}_{\text{tmp}}$ is time consuming. Then, we fine tune the network with our captured dataset. At this stage, $\lambda$, $\beta$, and $\gamma$ are set to 1, 0.5, 0.1 respectively. Note that, the temporal consistent loss is only applied to the denoising result in the raw domain since the temporal loss tends to smooth the image. Meanwhile, the reconstruction loss is applied to both the raw and sRGB denoising results. Although the parameters of the pretrained ISP are fixed before training the denoising network, this strategy is beneficial for improving the reconstruction quality in the sRGB domain.

\section{Experiments}



\begin{figure*}
    \centering
    \includegraphics[width=0.85\linewidth]{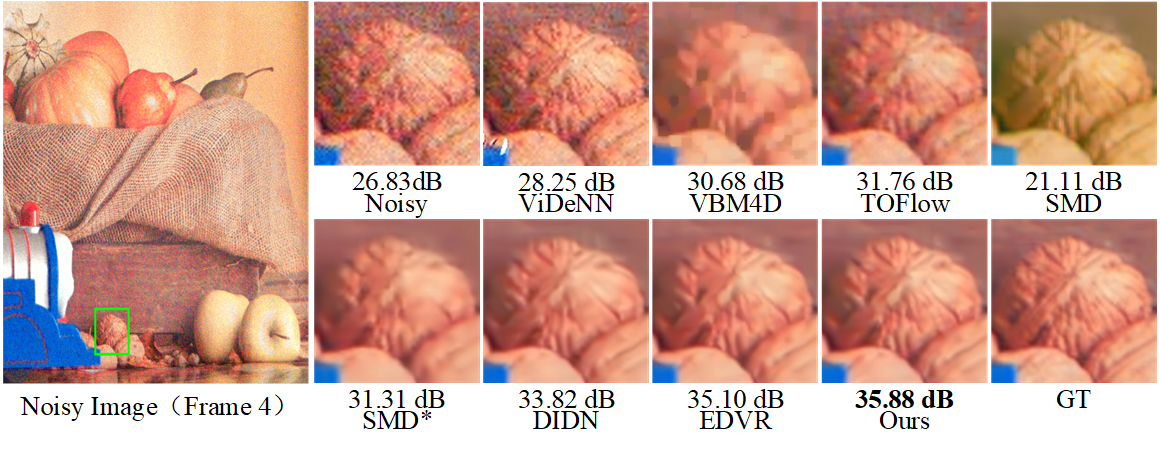}
    \caption{Visual quality comparison on one indoor scene captured under ISO 25600 (frame 4). Zoom in for better observation.}
    \label{fig:compare1}
\end{figure*}

\subsection{Training Details}
The channel number $C$ is set to 16 and the consecutive frame number $T$ is set to 3.
The size of the convolution filter size is $3\times 3$ and the upsampling process in pyramidal processing is realized by bilinear upsampling.
Our pre-denoising network is trained with learning rate 1e-4, and converges after 700 epochs. Our ISP network is pretrained with learning rate 1e-4, and converges after 770 epochs. The two networks are fixed during training the proposed RViDeNN.

We preprocess our synthetic and captured raw data by black level subtraction and white level normalization. Our network is trained with these processed raw data. During training, the patch size is set to $256\times 256$ (i.e. $H=W=128$ in the sub-sequences) and the batch size is set to 1. We first train our network using synthetic data with learning rate 1e-4. After 33 epochs, we finetune the network with our captured videos and the learning rate is set to 1e-6 except for the spatial fusion module, which is set to 1e-5. After 100 epochs, the whole network converges.
The proposed model is implemented in PyTorch and  trained with an NVIDIA 2080 TI GPU.

\subsection{Ablation Study}

In this section, we perform ablation study to demonstrate the effectiveness of the proposed raw domain processing, packing strategy for raw input, pre-denoising result guided alignment, and non-local attention modules in our network. Table \ref{Table:ablation} lists the quantitative comparison results in our captured test set by removing these modules one by one. It can be observed that the PSNR values in the sRGB domain is decreased by more than 1 dB compared with directly processing the noisy raw videos. By incorporating the packing strategy in raw denoising, i.e. processing the RGBG sub-sequences separately and merging them in the final stage, the denoising performance is nearly the same as that of the unpacking version. However, the parameters are greatly reduced since we only extract 16 channel features for each sub-sequence and the unpacking version extract 64 channel features. By further introducing the pre-denoising guided alignment module and non-local attention module, the PSNR values in the sRGB domain is improved by 0.26 dB.

\begin{table*}
\centering
\caption{Comparison with state-of-the-art denoising methods. Each row lists the average denoising results in raw (or sRGB) domain for 25 indoor videos. $\text{Ours}^{-}$ is the results generated by training the model with only synthetic dataset. The best results are highlighted in bold and the second best results are underlined.}
\resizebox{\textwidth}{11mm}{
\begin{tabular}{ccccccccccccccc}
\toprule
 \multicolumn{3}{c}{}                                                &Noisy& ViDeNN \cite{claus2019videnn} & VBM4D \cite{matteo2011video} & TOFlow \cite{xue2019video} & SMD \cite{chen2019seeing} & SMD* & EDVR \cite{wang2019edvr}        & DIDN \cite{yu2019deep}        &$\text{Ours}^{-}$ &   Ours       \\ \hline \hline

\multirow{4}{*}{} & \multicolumn{1}{l}{\multirow{2}{*}{Raw}}  & \multicolumn{1}{l}{PSNR} &32.01&-& -     & -      & -             & -            & -           & 43.25 &\underline{43.37}& \textbf{43.97} \\ \cline{3-13}
                         & \multicolumn{1}{l}{}                      & \multicolumn{1}{l}{SSIM} &0.732&-& -     & -      & -             & -            & -           & 0.984 &\underline{0.985}& \textbf{0.987} \\ \cline{2-13}
                         & \multicolumn{1}{l}{\multirow{2}{*}{sRGB}} & \multicolumn{1}{l}{PSNR} &31.79&31.48& 34.16 & 34.81  & 26.26         & 35.87        & 38.97 & 38.83  &\underline{39.19}   & \textbf{39.95} \\ \cline{3-13}
                         & \multicolumn{1}{l}{}                      & \multicolumn{1}{l}{SSIM} &0.752&0.826& 0.922 & 0.921  & 0.912         & 0.957        & 0.972       & 0.974 &\underline{0.975}& \textbf{0.979} \\
                         \bottomrule
\end{tabular}}
\label{table:results}
\end{table*}

%
%

\begin{table}
\centering
\caption{Ablation study for raw domain processing, packing, predenoising and non-local attention modules. The PSNR (or SSIM) results are the averaging results on all the testing videos under different ISO settings ranging from 1600 to 25600.}
\resizebox{0.47\textwidth}{15mm}{
\begin{tabular}{ccccccc}
\toprule
\multicolumn{2}{l}{Raw domain}   & $\times$  & $\checkmark$      & $\checkmark$    & $\checkmark$    & $\checkmark$     \\
\multicolumn{2}{l}{Packing}     & $\times$  & $\times$      & $\checkmark$    & $\checkmark$    & $\checkmark$     \\
\multicolumn{2}{l}{Pre-denoising}   & $\times$    &   $\times$    &  $\times$     & $\checkmark$    & $\checkmark$     \\
\multicolumn{2}{l}{Non-local attention}  &  $\times$   & $\times$      & $\times$      &  $\times$     & $\checkmark$     \\ \hline
\multirow{2}{*}{Raw}  & PSNR  & - & 43.84 & 43.84 & 43.88 & \textbf{43.97} \\
                      & SSIM   &  -  & 0.9866 & 0.9866 & 0.9871 & \textbf{0.9874} \\ \hline
\multirow{2}{*}{sRGB} & PSNR  & 38.58 & 39.69 & 39.69 & 39.80  & \textbf{39.95} \\
                      & SSIM   & 0.9703   & 0.9776 & 0.9778 & 0.9785 & \textbf{0.9792} \\
\bottomrule
\end{tabular}
}
\label{Table:ablation}
\end{table}

\subsection{Comparison with State-of-the-art Methods}
To demonstrate the effectiveness of the proposed denoising strategy, we compare with state-of-the-art video denoising methods, i.e. VBM4D \cite{matteo2011video}, TOFLow \cite{xue2019video}, ViDeNN \cite{claus2019videnn}, and SMD \cite{chen2019seeing}, video restoration method EDVR \cite{wang2019edvr}, and raw image denoising method DIDN \cite{yu2019deep}, which is the second winner of the NTIRE 2019 Challenge \cite{abdelhamed2019ntire} on real image denoising. We tune the noise level of VBM4D to generate the best denoising results. Since TOFLow and EDVR are designed for sRGB videos, we retrain the two networks using our sRGB noisy-clean video pairs. Since ViDeNN is a blind denoising method and there is no training code available, we directly utilize its released model. We give two results for SMD. The first result is generated with their pre-trained model and our raw image is preprocessed with their settings. In order to compare with our method in the full-resolution result, we did not utilize the binning process in SMD, and utilize the widely used demosaicing process \cite{brooks2018unprocessing} to preprocess our dataset for SMD. The second result is generated by retraining SMD (denoted as SMD*) with our dataset\footnote{Thanks to our multiple shots in generating the ground truth frame, we also have multiple noisy images for the same static scene.}. During retraining, we remove VBM4D pre-processing for a fair comparison. In the supplementary  material, we also give the retrained SMD results with VBM4D as pre-processing. DIDN is retrained with our noisy-clean image pairs, and its sRGB results are generated with our pre-trained ISP module. We evaluate these methods on 25 indoor testing videos with GT and 50 outdoor testing videos without GT.


Table \ref{table:results} lists the average denoising results for 25 indoor videos. Only DIDN and our method can produce both the raw and sRGB results. It can be observed that our method greatly outperforms the denoising methods conducted on sRGB domains. ViDeNN was not retrained with our dataset, and their pretrained model cannot handle the realistic noise captured under very high ISO values.  Since the original SMD is trained with a different dataset, its results have large colour cast, which leads to lower PSNR values. Compared with EDVR, which also utilizes alignment and fusion strategy, our method achieves nearly 1 dB gain. Compared with DIDN, our method achieves 0.72 dB gain in the raw domain and 1.12 dB gain in the sRGB domain. We also give our results generated by training with only synthetic dataset, denoted as $\text{Ours}^{-}$. $\text{Ours}^{-}$ still outperforms DIDN and EDVR. It demonstrates that our noise synthesis method is effective and the pretrained module is well generalized from high FPS outdoor scenes to low FPS indoor scenes.

\begin{figure}
    \centering
    \includegraphics[width=0.95\linewidth]{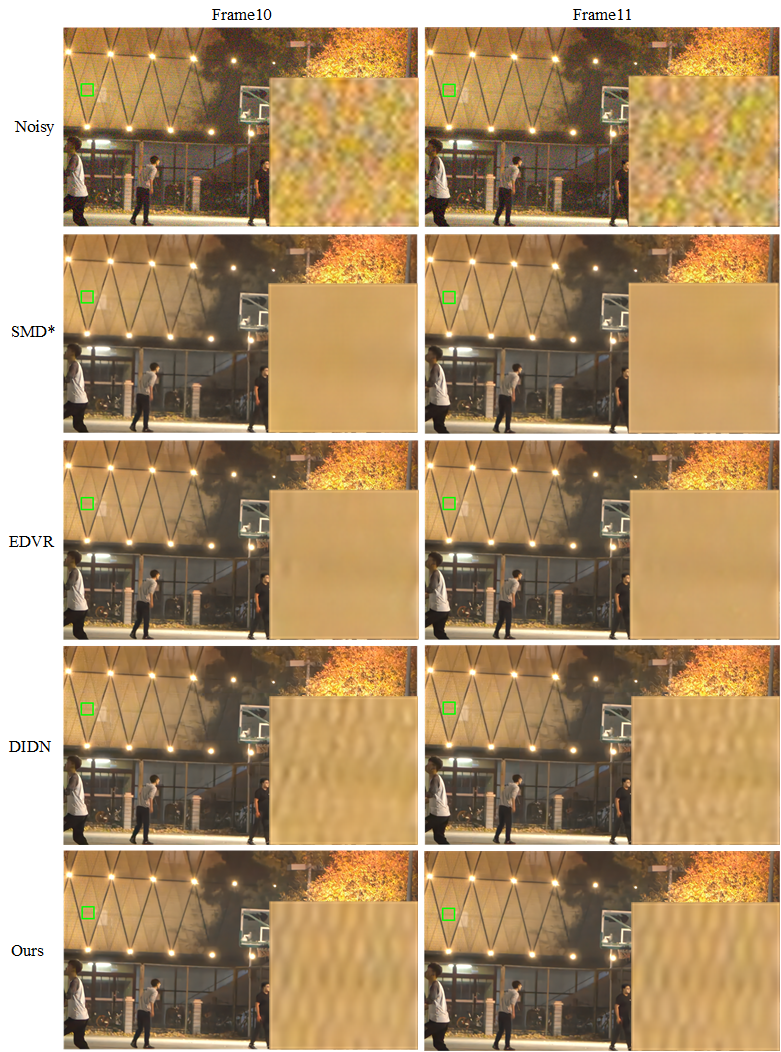}
    \caption{Visual quality comparison for two consecutive frames from one outdoor scene. Zoom in for better observation.}
    \label{fig:outdoor}
\end{figure}

Fig. \ref{fig:compare1}  presents the visual comparison results for one indoor scene captured under ISO 25600. It can be observed that our method removes the noise clearly and recovers the most fine-grained details. VBM4D, TOFlow and ViDeNN cannot remove the noise clearly. The results of SMD*, DIDN and EDVR are a bit smooth.
Fig. \ref{fig:outdoor} presents the outdoor denoising results. Due to page limits, we only present the comparison with SMD*, EDVR, and DIDN. It can be observed that the results of EDVR and DIDN are over-smooth. The recovered content is not consistent between neighboring frames for DIDN since it is a single image based denoising method. In contrast, our method removes the noise clearly and recovers temporal consistent textures.

Since there is no ground truth for the outdoor videos, we also conduct user study to evaluate the denoising performance for our outdoor dataset. The user study results and the demo for the video denoising results are provided in the supplementary material.

\section{Conclusion}
In this paper, we propose a RViDeNet by training on real noisy-clean video frames. By decomposing the raw sequences into RGBG sub-sequences and then going through the alignment, non-local attention, temporal fusion, and spatial fusion modules, our method fully takes advantage of the spatial, channel, and temporal correlations in the raw sequences. With both raw and sRGB outputs, our method gives users more flexibility in generating their favourite results. Experimental results demonstrate the superiority of the proposed method in removing realistic noise and producing temporally-consistent videos. We build the first noisy-clean dynamic video dataset, which will facilitate research on this topic.

\clearpage
{\small
\bibliographystyle{ieee_fullname}
\bibliography{egbib}

\begin{thebibliography}{10}\itemsep=-1pt

\bibitem{abdelhamed2018high}
Abdelrahman Abdelhamed, Stephen Lin, and Michael~S Brown.
\newblock A high-quality denoising dataset for smartphone cameras.
\newblock In {\em Proceedings of the IEEE Conference on Computer Vision and
  Pattern Recognition}, pages 1692--1700, 2018.

\bibitem{abdelhamed2019ntire}
Abdelrahman Abdelhamed, Radu Timofte, and Michael~S Brown.
\newblock Ntire 2019 challenge on real image denoising: Methods and results.
\newblock In {\em Proceedings of the IEEE Conference on Computer Vision and
  Pattern Recognition Workshops}, pages 0--0, 2019.

\bibitem{anaya2014renoir}
Josue Anaya and Adrian Barbu.
\newblock Renoir-a dataset for real low-light image noise reduction.
\newblock {\em arXiv preprint arXiv:1409.8230}, 2014.

\bibitem{anwar2019real}
Saeed Anwar and Nick Barnes.
\newblock Real image denoising with feature attention.
\newblock {\em Proceedings of International Conference on Computer Vision},
  2019.

\bibitem{brooks2018unprocessing}
Tim Brooks, Ben Mildenhall, Tianfan Xue, Jiawen Chen, Dillon Sharlet, and
  Jonathan~T Barron.
\newblock Unprocessing images for learned raw denoising.
\newblock {\em CVPR}, 2019.

\bibitem{buades2016patch}
Antoni Buades, Jose-Luis Lisani, and Marko Miladinovi{\'c}.
\newblock Patch-based video denoising with optical flow estimation.
\newblock {\em IEEE Transactions on Image Processing}, 25(6):2573--2586, 2016.

\bibitem{chen2019learning}
Chen Chen, Qifeng Chen, Minh~N. Do, and Vladlen Koltun.
\newblock Learning to see in the dark.
\newblock In {\em Proceedings of the IEEE Conference on Computer Vision and
  Pattern Recognition}, 2018.

\bibitem{chen2019seeing}
Chen Chen, Qifeng Chen, Minh~N. Do, and Vladlen Koltun.
\newblock Seeing motion in the dark.
\newblock In {\em Proceedings of the IEEE International Conference on Computer
  Vision}, 2019.

\bibitem{chen2018image}
Jingwen Chen, Jiawei Chen, Hongyang Chao, and Ming Yang.
\newblock Image blind denoising with generative adversarial network based noise
  modeling.
\newblock In {\em Proceedings of the IEEE Conference on Computer Vision and
  Pattern Recognition}, pages 3155--3164, 2018.

\bibitem{chen2016deep}
Xinyuan Chen, Li Song, and Xiaokang Yang.
\newblock Deep rnns for video denoising.
\newblock In {\em Applications of Digital Image Processing XXXIX}, volume 9971,
  page 99711T. International Society for Optics and Photonics, 2016.

\bibitem{claus2019videnn}
Michele Claus and Jan van Gemert.
\newblock Videnn: Deep blind video denoising.
\newblock In {\em Proceedings of the IEEE Conference on Computer Vision and
  Pattern Recognition Workshops}, pages 0--0, 2019.

\bibitem{dabov2007image}
Kostadin Dabov, Alessandro Foi, Vladimir Katkovnik, and Karen Egiazarian.
\newblock Image denoising by sparse 3-d transform-domain collaborative
  filtering.
\newblock {\em IEEE Transactions on image processing}, 16(8):2080--2095, 2007.

\bibitem{dai2017deformable}
Jifeng Dai, Haozhi Qi, Yuwen Xiong, Yi Li, Guodong Zhang, Han Hu, and Yichen
  Wei.
\newblock Deformable convolutional networks.
\newblock In {\em Proceedings of the IEEE international conference on computer
  vision}, pages 764--773, 2017.

\bibitem{FoiNoise}
Alessandro Foi, Sakari Alenius, Vladimir Katkovnik, and Karen Egiazarian.
\newblock Noise measurement for raw-data of digital imaging sensors by
  automatic segmentation of nonuniform targets.
\newblock {\em IEEE Sensors Journal}, 7(10):1456--1461.

\bibitem{FoiPractical}
A. Foi, M. Trimeche, V. Katkovnik, and K. Egiazarian.
\newblock Practical poissonian-gaussian noise modeling and fitting for
  single-image raw-data.
\newblock 17(10):1737--1754.

\bibitem{fu2019dual}
Jun Fu, Jing Liu, Haijie Tian, Yong Li, Yongjun Bao, Zhiwei Fang, and Hanqing
  Lu.
\newblock Dual attention network for scene segmentation.
\newblock In {\em Proceedings of the IEEE Conference on Computer Vision and
  Pattern Recognition}, pages 3146--3154, 2019.

\bibitem{gharbi2016deep}
Micha{\"e}l Gharbi, Gaurav Chaurasia, Sylvain Paris, and Fr{\'e}do Durand.
\newblock Deep joint demosaicking and denoising.
\newblock {\em ACM Transactions on Graphics (TOG)}, 35(6):191, 2016.

\bibitem{godard2018deep}
Cl{\'e}ment Godard, Kevin Matzen, and Matt Uyttendaele.
\newblock Deep burst denoising.
\newblock In {\em Proceedings of the European Conference on Computer Vision
  (ECCV)}, pages 538--554, 2018.

\bibitem{guo2019toward}
Shi Guo, Zifei Yan, Kai Zhang, Wangmeng Zuo, and Lei Zhang.
\newblock Toward convolutional blind denoising of real photographs.
\newblock In {\em Proceedings of the IEEE Conference on Computer Vision and
  Pattern Recognition}, pages 1712--1722, 2019.

\bibitem{huang2019ccnet}
Zilong Huang, Xinggang Wang, Lichao Huang, Chang Huang, Yunchao Wei, and Wenyu
  Liu.
\newblock Ccnet: Criss-cross attention for semantic segmentation.
\newblock In {\em Proceedings of the IEEE International Conference on Computer
  Vision}, pages 603--612, 2019.

\bibitem{ji2010robust}
Hui Ji, Chaoqiang Liu, Zuowei Shen, and Yuhong Xu.
\newblock Robust video denoising using low rank matrix completion.
\newblock In {\em 2010 IEEE Computer Society Conference on Computer Vision and
  Pattern Recognition}, pages 1791--1798. IEEE, 2010.

\bibitem{liang2019cameranet}
Zhetong Liang, Jianrui Cai, Zisheng Cao, and Lei Zhang.
\newblock Cameranet: A two-stage framework for effective camera isp learning.
\newblock {\em arXiv preprint arXiv:1908.01481}, 2019.

\bibitem{liu2019learning}
Jiaming Liu, Chi-Hao Wu, Yuzhi Wang, Qin Xu, Yuqian Zhou, Haibin Huang, Chuan
  Wang, Shaofan Cai, Yifan Ding, Haoqiang Fan, et~al.
\newblock Learning raw image denoising with bayer pattern unification and bayer
  preserving augmentation.
\newblock In {\em Proceedings of the IEEE Conference on Computer Vision and
  Pattern Recognition Workshops}, pages 0--0, 2019.

\bibitem{matteo2011video}
Maggioni Matteo, Giacomo Boracchi, Foi Alessandro, Egiazarian Karen, et~al.
\newblock Video denoising using separable 4d nonlocal spatiotemporal
  transforms.
\newblock In {\em Image Processing: Algorithms and Systems IX}, pages 1--11.
  SPIE, 2011.

\bibitem{MilanMOT16}
Anton Milan, Laura Leal-Taixe, Ian Reid, Stefan Roth, and Konrad Schindler.
\newblock Mot16: A benchmark for multi-object tracking.

\bibitem{MildenhallBurst}
Ben Mildenhall, Jonathan~T. Barron, Jiawen Chen, Dillon Sharlet, and Robert
  Carroll.
\newblock Burst denoising with kernel prediction networks.

\bibitem{mildenhall2018burst}
Ben Mildenhall, Jonathan~T Barron, Jiawen Chen, Dillon Sharlet, Ren Ng, and
  Robert Carroll.
\newblock Burst denoising with kernel prediction networks.
\newblock In {\em Proceedings of the IEEE Conference on Computer Vision and
  Pattern Recognition}, pages 2502--2510, 2018.

\bibitem{nam2016holistic}
Seonghyeon Nam, Youngbae Hwang, Yasuyuki Matsushita, and Seon Joo~Kim.
\newblock A holistic approach to cross-channel image noise modeling and its
  application to image denoising.
\newblock In {\em Proceedings of the IEEE Conference on Computer Vision and
  Pattern Recognition}, pages 1683--1691, 2016.

\bibitem{plotz2017benchmarking}
Tobias Plotz and Stefan Roth.
\newblock Benchmarking denoising algorithms with real photographs.
\newblock In {\em Proceedings of the IEEE Conference on Computer Vision and
  Pattern Recognition}, pages 1586--1595, 2017.

\bibitem{ratnasingam2019deep}
Sivalogeswaran Ratnasingam.
\newblock Deep camera: A fully convolutional neural network for image signal
  processing.
\newblock In {\em Proceedings of the IEEE International Conference on Computer
  Vision Workshops}, pages 0--0, 2019.

\bibitem{ronneberger2015u}
Olaf Ronneberger, Philipp Fischer, and Thomas Brox.
\newblock U-net: Convolutional networks for biomedical image segmentation.
\newblock In {\em International Conference on Medical image computing and
  computer-assisted intervention}, pages 234--241. Springer, 2015.

\bibitem{schwartz2018deepisp}
Eli Schwartz, Raja Giryes, and Alex~M Bronstein.
\newblock Deepisp: learning end-to-end image processing pipeline.
\newblock {\em arXiv preprint arXiv:1801.06724}, 2018.

\bibitem{tassano2019dvdnet}
Matias Tassano, Julie Delon, and Thomas Veit.
\newblock Dvdnet: A fast network for deep video denoising.
\newblock 2019.

\bibitem{tassano2019fastdvdnet}
Matias Tassano, Julie Delon, and Thomas Veit.
\newblock Fastdvdnet: Towards real-time video denoising without explicit motion
  estimation.
\newblock {\em arXiv preprint arXiv:1907.01361}, 2019.

\bibitem{wang2019edvr}
Xintao Wang, Kelvin~CK Chan, Ke Yu, Chao Dong, and Chen Change~Loy.
\newblock Edvr: Video restoration with enhanced deformable convolutional
  networks.
\newblock In {\em Proceedings of the IEEE Conference on Computer Vision and
  Pattern Recognition Workshops}, pages 0--0, 2019.

\bibitem{wang2018non}
Xiaolong Wang, Ross Girshick, Abhinav Gupta, and Kaiming He.
\newblock Non-local neural networks.
\newblock In {\em Proceedings of the IEEE conference on computer vision and
  pattern recognition}, pages 7794--7803, 2018.

\bibitem{woocbam}
Sanghyun Woo, Jongchan Park, Joon-Young Lee, and In~So Kweon.
\newblock Cbam: Convolutional block attention module.
\newblock In {\em Europe Conference on Computer Vision}, 2018.

\bibitem{xu2018real}
Jun Xu, Hui Li, Zhetong Liang, David Zhang, and Lei Zhang.
\newblock Real-world noisy image denoising: A new benchmark.
\newblock {\em arXiv preprint arXiv:1804.02603}, 2018.

\bibitem{xu2018external}
Jun Xu, Lei Zhang, and David Zhang.
\newblock External prior guided internal prior learning for real-world noisy
  image denoising.
\newblock {\em IEEE Transactions on Image Processing}, 27(6):2996--3010, 2018.

\bibitem{xu2018trilateral}
Jun Xu, Lei Zhang, and David Zhang.
\newblock A trilateral weighted sparse coding scheme for real-world image
  denoising.
\newblock {\em ECCV}, 2018.

\bibitem{xu2017multi}
Jun Xu, Lei Zhang, David Zhang, and Xiangchu Feng.
\newblock Multi-channel weighted nuclear norm minimization for real color image
  denoising.
\newblock In {\em IEEE International Conference on Computer Vision}, volume~2,
  2017.

\bibitem{xu2019towards}
Xiangyu Xu, Yongrui Ma, and Wenxiu Sun.
\newblock Towards real scene super-resolution with raw images.
\newblock In {\em Proceedings of the IEEE Conference on Computer Vision and
  Pattern Recognition}, pages 1723--1731, 2019.

\bibitem{xue2019video}
Tianfan Xue, Baian Chen, Jiajun Wu, Donglai Wei, and William~T Freeman.
\newblock Video enhancement with task-oriented flow.
\newblock {\em International Journal of Computer Vision}, 127(8):1106--1125,
  2019.

\bibitem{yu2019deep}
Songhyun Yu, Bumjun Park, and Jechang Jeong.
\newblock Deep iterative down-up cnn for image denoising.
\newblock In {\em Proceedings of the IEEE Conference on Computer Vision and
  Pattern Recognition Workshops}, pages 0--0, 2019.

\bibitem{yue2019high}
Huanjing Yue, Jianjun Liu, Jingyu Yang, Truong Nguyen, and Feng Wu.
\newblock High iso jpeg image denoising by deep fusion of collaborative and
  convolutional filtering.
\newblock {\em IEEE Transactions on Image Processing}, 2019.

\bibitem{zhang2019zoom}
Xuaner Zhang, Qifeng Chen, Ren Ng, and Vladlen Koltun.
\newblock Zoom to learn, learn to zoom.
\newblock In {\em Proceedings of the IEEE Conference on Computer Vision and
  Pattern Recognition}, pages 3762--3770, 2019.

\end{thebibliography}
}

\end{document}